\documentstyle[12pt,epsf]{article}
\newcommand{\includegraphics}[1]{\leavevmode\epsffile{#1}}

\begin{document}

\title{Dynamical Perturbation Theory\\and the Auxiliary Field
Method}
\author{B. Holdom\footnote{email: holdom@utcc.utoronto.ca}\\
\small\em Department of Physics, University of
Toronto\\\small\em Toronto ON M5S1A7, Canada  \and
Qing Wang\footnote{email: wangqing@bepc2.ihep.ac.cn}\\
\small\em China Center of Advanced Science and Technology (World
Laboratory)\\\small\em  P.O.Box 8730, Beijing 100080, China \\
\small\em Institute of Modern Physics, Tsinghua University, Beijing 100084,
China
\footnote{Mailing address}\\
\small\em Institute of Theoretical Physics, Academia Sinica, Beijing 100080,
China\\\small\em  Department of Physics, Tsinghua University, Beijing 100084,
China}
\date{}
\maketitle
\begin{abstract}
By introducing auxiliary fields, which by their quantum numbers vanish
in perturbation theory, we relate the dynamical
perturbation theory of Pagels and Stokar and a successful gauged
nonlocal constituent quark model to a $U(N)$ gauge theory and to QCD.
This sheds light on the duality between quark models and resonance
models. We then derive the effective action in the loop expansion
without dropping cubic and quartic gluon couplings.
\end{abstract}

\section{Introduction} Consider the full fermion propagator
in the presence of a dynamically broken chiral symmetry in a gauge
theory with no explicit fermion masses.
\begin{equation}{S}^{-1}(p)=Z({p}^{2}){p}\!\!\!{/}-\Sigma({p}^{2})
\label{a}\end{equation}
$\Sigma({p}^{2})$ vanishes to all orders in perturbation theory and
may be considered to have the dependence ${e}^{-1/b{g}^{2}}$ on the
gauge coupling $g$ and some constant $b$. To account for this
dynamical momentum-dependent mass function, Pagels and Stokar
introduced ``dynamical perturbation theory" (DPT)
\cite{ps}. We quote from their work. \begin{quote} To lowest order
this approximation can be described as follows. Amplitudes that do
not vanish to all orders of perturbation theory are given by their
free-field values. For example, the amplitude
$Z(p^2)$ in (\ref{a}) is given by
$Z({p}^{2})=1$. Amplitudes that vanish in all orders in perturbation
theory such as
$\lambda ={e}^{-1/b{g}^{2}}$, for example $\Sigma({p}^{2})$, are
retained, but we drop terms of order ${g}^{n}{e}^{-1/b{g}^{2}}$,
$n>0$. The virtue of this approximation scheme is that amplitudes can
be made to satisfy the chiral Ward identities consistently to lowest
order in the parameters $g^2$ and $\lambda
={e}^{-1/b{g}^{2}}$.\end{quote} One result of this work is a popular
formula relating the (pseudo)Goldstone boson decay constant to the
fermion mass function, which has been widely used in studies of QCD
and other theories with chiral symmetry breaking. This formula is
found to be in good agreement with a more sophisticated
Bethe-Salpeter approach \cite{aoki}.

Since the Ward identities are satisfied the idea of DPT at lowest
order can be extended to a Lagrangian-based GNC (gauged nonlocal
constituent) relativistic quark model \cite{gnc} which preserves the
chiral structure of QCD. The Lagrangian must be nonlocal to
incorporate the momentum dependent mass. The GNC model reproduced the
Pagels-Stokar decay constant formula, and when expanded to
order $p^4$ in the energy expansion was found to be surprisingly
successful at describing low energy chiral dynamics. The
momentum dependence of the quark mass function provides a natural
ultraviolet cutoff on the quark loop integrals, thus leading to the
correct Wess-Zumino terms without regularization subtleties
\cite{ball}. The incorporation of this momentum dependence
represents an improvement over constant mass quark models of the
Nambu-Jona-Lasino type. But the question remains, why does a free
quark model represent low energy QCD dynamics as well as it does? For
example at the level of the quark propagator, how can the momentum
dependence of
$Z(p^2)$ in (\ref{a}) be ignored?

By using the auxiliary field method we shall provide a systematic
derivation of DPT and the GNC quark model from a gauge theory. The
connection with DPT is made by introducing auxiliary fields only for
those quark bilinears which define purely nonperturbative amplitudes.
The auxiliary fields introduced are thus chirality changing, and they
may also carry color or fermion number. The effective theory at
lowest order in a loop expansion will correspond to the lowest order
in DPT, and the origin of the GNC quark model is made clear.

We shall
also find a set of ladder SD equations which are all homogeneous
equations in the auxiliary fields. It may be noted
that these SD equations treat all the possible quark mass functions
in the various channels, including $\Sigma(p)$ for the usual
$\overline{q}q$ mass, on an equal footing. In the SD equation for
$\Sigma(p)$ we will automatically have
$Z(p)=1$, which is usually obtained in studies using the ladder SD
equation via a special choice of gauge. In our approach
$Z(p)=1$ follows at lowest order in the loop expansion.

We shall deal with the problem of gauge boson self-interactions by
studying a strongly interacting $U(N_c)$ (rather than $SU(N_c)$) gauge
theory with
$N_f$ flavors of quarks. In $U(N_c)$ there are two coupling constants,
$g$ and
$g_1$ for $SU(N_c)$ and $U(1)$ respectively. Both of these interactions
are attractive in the usual color- and flavor-singlet $\overline{q}q$
channel.  For chiral symmetry breaking to occur in this channel there
is some constraint on some combination of $g$ and $g_1$. We shall
take advantage of the fact that we can hold the scale of chiral
symmetry breaking fixed while
$g$ is free to vary. In this way the theory can smoothly interpolate
between the QCD limit (large $g$ and small $g_1$) and the $U(1)$
limit (small $g$ and large
$g_1$). For some subset of this space of theories it makes sense to
treat
$g$ perturbatively on the momentum scales which characterize chiral
symmetry breaking. For some scale sufficiently below the chiral
symmetry breaking scale
$g$ will grow strong and cause confinement. But the effects due to
the nonabelian nature of $SU(N_c)$, namely trilinear and quartic gauge
couplings, may be treated perturbatively as far as the study of
chiral symmetry breaking is concerned.

Our approach will also shed some light on the concept of duality in
the description of low energy chiral dynamics. This is related to
the observation that the low energy chiral Lagrangian to order $p^4$
of QCD is surprisingly well described by constituent quark models,
such as the GNC quark model. The quark degrees of freedom are somehow
offering a dual description to the effects of the lowest lying vector
and axial vector resonances, which are usually thought to dominate
low energy chiral dynamics. We shed light on this by showing that QCD
may be rewritten as a path integral over various auxiliary field
degrees of freedom which {\bf do not} include  vector or axial vector
degrees of freedom. Instead we will have scalar and tensor fields,
some of which have color and/or diquark quantum numbers. The lowest
order in a loop expansion involving all massive degrees of freedom
corresponds to setting all the non-Goldstone-boson degrees of freedom
in the various auxiliary fields to their vacuum values. The fact that
we are not required to fix vector or axial-vector degrees of freedom
then suggests how it is that a constituent quark model can reproduce
the effects of vector or axial-vector degrees of freedom.

Our basic result is that the perturbative expansion of a nonabelian
gauge theory may be reorganized in such a way that the lowest order
in the new loop expansion coincides with a previously studied and
successful quark model of low energy chiral dynamics. Instead of
being put in by hand, the quark mass function emerges from the
dynamics. Also the quark model now exists in a framework of a loop
expansion in which corrections, at least in principle, may be
systematically worked out. Such a process was implied but not
specified in the original proposal for DTP. As one example the
phenomenological effects of various scalar and diquark degrees of
freedom could be explored in the context of a quark model, without
fear of double counting degrees of freedom.

In the next two sections we shall derive generating functionals
without and with, respectively, the effects of cubic and quartic
gauge couplings. The latter case is significantly more complicated,
and it will be carried out by introducing two sets of auxiliary
fields. In the next section we also show how the (pseudo)Goldstone
degrees of freedom may be retained for a description of low energy
physics.

\section{Low energy theory} Due to our study of $U(N_c)$ rather than
$SU(N_c)$, we will first ignore effects from the cubic and quartic gauge
couplings from
the beginning. Then the gauge fields may be integrated out to give
the following generating functional.
\begin{eqnarray} e^{iW[J]}&=&\int{\cal D}\psi{\cal
D}\overline{\psi}{\rm exp}i{\bigg\{}{\int}d^{4}x
\overline{\psi}(i\partial\!\!\!/+J){\psi}\nonumber\\
 &&+\frac{i^2g^2}{2}
 \int d^4x d^4 yG_{\mu\nu}(x,y)[\overline{\psi}^i_{\alpha}(x)
 (\frac{\lambda_a}{2})_{\alpha\beta}\gamma^{\mu}{\psi}^i_{\beta}(x)]
[\overline{\psi}^j_{\alpha'}(y)(\frac{\lambda_a}{2})_{\alpha'\beta'}
\gamma^{\nu}{\psi}^j_{\beta'}(y)]\nonumber\\ 
 &&+\frac{i^2g^2_1}{2}
 \int d^4x d^4 yG_{\mu\nu}^{(1)}(x,y)[\overline{\psi}^i_{\alpha}(x)
 \gamma^{\mu}{\psi}^i_{\alpha}(x)]
 [\overline{\psi}^j_{\alpha'}(y)\gamma^{\nu}{\psi}^j_{\alpha'}(y)]
 \bigg\}
\end{eqnarray}
$G_{\mu\nu}(x,y)$ and
$G_{\mu\nu}^{(1)}(x,y)$ are the tree level $SU(N_c)$ and $U(1)$ gauge
field propagators respectively. 
$J$ represents the external sources which are nontrivial matrices in
the $SU(N_f)$ flavor space.
\begin{equation} J(x)=v\!\!\! /\;(x)+a\!\!\!
/\;(x)\gamma_5-s(x)+ip(x)\gamma_5\nonumber
\end{equation} The generating functional may be rearranged as
follows. 
\begin{eqnarray} e^{iW[J]}&=&\int{\cal D}\psi{\cal
D}\overline{\psi}{\rm exp}i{\bigg\{}{\int}d^{4}x
\overline{\psi}(i\partial\!\!\!/+J){\psi}+\frac{i^2g^2}{2}
 \int d^4x d^4 yG_{\mu\nu}(x,y)\bigg[\nonumber\\ &&
2[\overline{\psi}^i_{R,\alpha}(x)
(\frac{\lambda_a}{2})_{\alpha\beta}\gamma^{\mu}{\psi}^i_{R,\beta}(x)]
[\overline{\psi}^j_{L,\alpha'}(y)(\frac{\lambda_a}{2})_{\alpha'\beta'}
\gamma^{\nu}{\psi}^j_{L,\beta'}(y)]\bigg]\nonumber\\ &&+
[\overline{\psi}^i_{L,\alpha}(x)
(\frac{\lambda_a}{2})_{\alpha\beta}\gamma^{\mu}{\psi}^i_{L,\beta}(x)]
[\overline{\psi}^j_{L,\alpha'}(y)(\frac{\lambda_a}{2})_{\alpha'\beta'}
\gamma^{\nu}{\psi}^j_{L,\beta'}(y)]\nonumber\\ &&+
[\overline{\psi}^i_{R,\alpha}(x)
(\frac{\lambda_a}{2})_{\alpha\beta}\gamma^{\mu}{\psi}^i_{R,\beta}(x)]
[\overline{\psi}^j_{R,\alpha'}(y)(\frac{\lambda_a}{2})_{\alpha'\beta'}
\gamma^{\nu}{\psi}^j_{R,\beta'}(y)]\nonumber\\
&&+\frac{i^2g^2_1}{2}
 \int d^4x d^4 yG_{\mu\nu}^{(1)}(x,y)\bigg[
2[\overline{\psi}^i_{R,\alpha}(x)\gamma^{\mu}{\psi}^i_{R,\alpha}(x)]
[\overline{\psi}^j_{L,\alpha'}(y)\gamma^{\nu}{\psi}^j_{L,\alpha'}(y)]
\nonumber\\ &&+
 [\overline{\psi}^i_{L,\alpha}(x)\gamma^{\mu}{\psi}^i_{L,\alpha}(x)]
[\overline{\psi}^j_{L,\alpha'}(y)\gamma^{\nu}{\psi}^j_{L,\alpha'}(y)]
+
[\overline{\psi}^i_{R,\alpha}(x)\gamma^{\mu}{\psi}^i_{R,\alpha}(x)]
[\overline{\psi}^j_{R,\alpha'}(y)\gamma^{\nu}{\psi}^j_{R,\alpha'}(y)]
 \bigg]\bigg\}\nonumber
\\ &=&\int_{}^{}{\cal D}\psi {\cal
D}\overline{\psi }{\rm exp}i\left\{{\int_{}^{}{d}^{4}x\overline{\psi
}(i{\partial }\!\!\!{/}+J)\psi
}\right.\nonumber\\
&&-\left.{{\frac{1}{2}}\int_{}^{}{d}^{4}x{d}^{4}y}\right.\left[{2{\overline{
\psi
}}_{R,\alpha }^{i\sigma }(x){\psi }_{L,{\beta }^{\prime }}^{j{\rho
}^{\prime }}(y)K(x,y{)}_{\alpha \beta {\alpha }^{\prime }{\beta
}^{\prime }}^{\sigma \rho {\sigma }^{\prime }{\rho }^{\prime
}}{\overline{\psi }}_{L,{\alpha }^{\prime }}^{j{\sigma }^{\prime
}}(y){\psi }_{R,\beta }^{i\rho }(x)}\right.\nonumber\\ 
&&\left.{\left.{[-{\overline{\psi }}_{L,\alpha }^{i\sigma
}(x){\overline{\psi }}_{L,{\alpha }^{\prime }}^{j{\sigma }^{\prime
}}(y)K(x,y{)}_{\alpha \beta {\alpha }^{\prime }{\beta }^{\prime
}}^{\sigma \rho {\sigma }^{\prime }{\rho }^{\prime }}{\psi
}_{L,{\beta }^{\prime }}^{j{\rho }^{\prime }}(y){\psi }_{L,\beta
}^{i\rho }(x)-(L\rightarrow R)]}\right]}\right\}\label{b}
\end{eqnarray} where
\begin{equation} K(x,y{)}_{\alpha \beta  {\alpha }^{\prime }{\beta 
}^{\prime }}^{\sigma
\rho {\sigma }^{\prime }{\rho }^{\prime }} ={g}^{2}{G}_{\mu \nu
}(x,y)({\frac{{\lambda }_{a}}{2}}{)}_{\alpha \beta  }({\gamma }^{\mu
}{)}_{\sigma \rho })({\frac{{\lambda }_{a}}{2}}{)}_{{\alpha }^{\prime
}{\beta  }^{\prime }}({\gamma }^{\mu }{)}_{{\sigma }^{\prime }{\rho
}^{\prime }} +{g}_{1}^{2}{G}_{\mu \nu }^{(1)}{\delta }_{\alpha \beta 
}({\gamma }^{\mu }{)}_{\sigma \rho }){\delta }_{{\alpha }^{\prime
}{\beta  }^{\prime }}({\gamma }^{\mu }{)}_{{\sigma }^{\prime }{\rho
}^{\prime }}\end{equation}

We may follow the auxiliary field method \cite{kugo} and insert a
Gaussian integral to cancel the four-fermion terms. Note that our
choice of auxiliary fields differs from previous treatments.
\begin{eqnarray}{\rm constant}&=&\int_{}^{}{\cal D}{\chi }_{R}{\cal
D}{\chi }_{L}{\cal D}{\kappa }_{L}{\cal D}{\kappa }_{R}{\cal
D}{\overline{\kappa }}_{L}{\cal D}{\overline{\kappa }}_{R}{\rm
exp}i\int_{}^{}{d}^{4}x{d}^{4}y\nonumber\\ &&\times
\left\{{{\rm Tr}\left[{\left({{\chi }_{R}-K{\overline{\psi
}}_{R}{\psi }_{L}}\right){K}^{-1}\left({{\chi }_{L}-K{\overline{\psi
}}_{L}{\psi }_{R}}\right)}\right]}\right.\nonumber\\ &&\left.{-
{\frac{1}{2}}{\rm Tr}\left[{\left({{\overline{\kappa
}}_{L}-K{\overline{\psi }}_{L}{\overline{\psi
}}_{L}}\right){K}^{-1}\left({{\kappa }_{L}-K{\psi }_{L}{\psi
}_{L}}\right)+ (L\rightarrow R)}\right] }\right\}
\end{eqnarray} The various auxiliary fields are bilocal, ${\chi
}_{L}(x,y)$, etc., and the traces imply contraction of indices
similar to (\ref{b}). We will not need the explicit form of
${K}^{-1}$. The resulting action takes the form
\begin{eqnarray}{e}^{iW[J]}&=&\int_{}^{}{\cal D}{\chi }_{R}{\cal
D}{\chi }_{L}{\cal D}{\kappa }_{L}{\cal D}{\kappa }_{R}{\cal
D}{\overline{\kappa }}_{L}{\cal D}{\overline{\kappa }}_{R}{\cal
D}\psi {\cal D}\overline{\psi }{\rm
exp}i\left\{{\int_{}^{}{d}^{4}x\overline{\psi }(i{\partial
}\!\!\!{/}+J)\psi }\right.\nonumber\\
&&+\int_{}^{}\left.{d}^{4}x{d}^{4}y\right[{\rm
Tr}\left({{\chi }_{R}{K}^{-1}{\chi }_{L}}\right)-{\frac{1}{2}}{\rm
Tr}\left({{\overline{\kappa }}_{L}{K}^{-1}{\kappa
}_{L}}\right)-{\frac{1}{2}}{\rm Tr}\left({{\overline{\kappa
}}_{R}{K}^{-1}{\kappa }_{R}}\right)\nonumber\\
&&\left.{\left.{-{\overline{\psi }}_{L}{\chi }_{R}{\psi
}_{R}-{\overline{\psi }}_{R}{\chi }_{L}{\psi }_{L}-{\psi
}_{L}{\overline{\kappa }}_{L}{\psi }_{L}-{\overline{\psi
}}_{L}{\kappa }_{L}{\overline{\psi }}_{L}-{\psi
}_{R}{\overline{\kappa }}_{R}{\psi }_{R}-{\overline{\psi
}}_{R}{\kappa }_{R}{\overline{\psi }}_{R}}\right]}\right\}\label{c}
\end{eqnarray}

The fermion fields could now be integrated out. The original path
integral over the quark and gluon fields has been traded in for a
path integral over various bosonic degrees of freedom. We first consider
the
lowest order in the loop expansion with respect to all auxiliary
fields.  The stationary condition for the tree level action with
$J=0$ gives a set of SD equations, homogeneous in the auxiliary
fields. The auxiliary fields may be decomposed into various
irreducible representations of the $SU({N}_{c})\times SU({N}_{f})$
color and flavor symmetries. If we define \begin{equation}\chi\equiv
{\chi }_{L}+{\chi }_{R}\end{equation} then the SD equation for the
flavor- and color-singlet degree of freedom, ${\chi (x,y)}_{\alpha
\beta ,ij}^{\rho
\sigma }\rightarrow{\delta }_{\alpha \beta }{\delta }_{\rho \sigma
}{\delta }_{ij}\Sigma (x-y)$,
is
\begin{eqnarray}\Sigma
(x-y)&=&-\left[{{g}^{2}\left({{\frac{{N}_{c}^{2}-1}{{N}_{c}}}}\right)
{G}_{\mu
\nu }(x-y)+{g}_{1}^{2}{G}_{\mu \nu }^{(1)}(x-y)}\right]\nonumber\\
&&\times
{{\gamma }^{\mu
}\left[{{P}_{L}S(x-y){P}_{L}+{P}_{R}S(x-y){P}_{R}}\right]}{\gamma
}^{\nu },\end{eqnarray}
\begin{equation}{\rm with}\,\,\,\,\,\,i{S}^{-1}(x-y)=i{\partial
}\!\!\!{/}\delta (x-y)-\Sigma (x-y).\end{equation} In the presence
of an ultraviolet cutoff this equation will determine a mass function
$\Sigma (p)$. All other auxiliary fields are in less attractive
channels and we may assume that they take on the vanishing solutions.

We note that
except for the (pseudo)Goldstone bosons (PGBs) associated with chiral
symmetry breaking, all other auxiliary field degrees of freedom are
expected to have masses of order the chiral symmetry breaking scale
or higher. This leads us to consider the lowest order in a loop
expansion with respect to all massive, non-Goldstone degrees of
freedom. That
is, we freeze all massive auxiliary fields to their vacuum ($J=0$)
values as determined by the SD equations above, but we wish to retain
the light PGB degrees of freedom as dynamical fields in a low energy
theory. Note that nonzero sources
$J$ are needed to define the generating functional of the low energy
theory. Nonzero sources are consistent with the freezing of the
massive fields at their $J=0$ vacuum values as long as the sources
only vary on distance scales large compared with inverse masses of the
massive degrees of freedom. We will say more about the effects of
the massive degrees of freedom at the end of this section.

We constrain the massive degrees of freedom as follows.
\begin{eqnarray}{{\rm tr}}_{c}[{\lambda }_{a}\chi
(x,y)]&=&0\\
{\rm tr}[{\sigma }_{\mu \nu }\chi
(x,y)]&=&0\\
\sigma (x,y{)}_{ik}\sigma (y,x{)}_{kj}+\Pi
(x,y{)}_{ik}\Pi (y,x{)}_{kj}&=&{\delta }_{ij}\Sigma
(x-y{)}^{2}\label{d}\\
\sigma ={\rm tr}(\chi
)/4,\,\,\Pi &=&{\rm tr}(i{\gamma }_{5}\chi )/4\end{eqnarray}
tr (${\rm
tr}_c$) denotes a trace over spinor (color) indices only. The first
constraint keeps $\chi (x,y)$ diagonal in color space, and we
henceforth ignore color indices. The second constraint sets the
tensor degrees of freedom to zero. The third constraint, where $i$,
$j$ are flavor indices, allows for PGB degrees of freedom in flavor
space while constraining the scalar mode.

With these constraints on the fields we may now go back and consider
the
${K}^{-1}$ terms in (\ref{c}). These terms are quadratic in $\sigma$
and $\pi$ and by chiral symmetry they must take the
form
\begin{equation}\int_{}^{}{d}^{4}x{d}^{4}y\left[{{\sigma
}_{ij}(x,y){\sigma }_{ji}(y,x)+{\Pi }_{ij}(x,y){\Pi
}_{ji}(y,x)}\right]f(x-y)\end{equation} The constraint (\ref{d})
then implies that these terms are independent of the PGB fields. The
PGB degrees of freedom are then described completely by the Trln
terms, and the generating functional for slowly varying sources is
\begin{eqnarray}{e}^{iW[J]}&=&\int_{}^{}{\cal D}U{\rm
exp}\left[{-i{\rm Trln}\left({i{\partial }\!\!\!{/}+J+{\chi }_{\pi
,J}(x,y)}\right)}\right]\\
U(x)&\equiv& {e}^{2i\pi (x){\gamma }_{5}/{f}_{\pi
}}\end{eqnarray}

${\chi }_{\pi ,J}(x,y)$ must be chosen to satisfy the constraints
above and to transform correctly under chiral transformations
of $\pi(x)$ and $J(x)=v\!\!\! /\;(x)+a\!\!\!
/\;(x)\gamma_5-s(x)+ip(x)\gamma_5$. A choice for ${\chi }_{\pi ,J}(x,y)$
corresponds to a choice
of representation of the $\pi(x)$ fields. A minimal choice is the
following \cite{gnc}.
\begin{eqnarray}&&{\chi }_{\pi ,J}(x,y)=\Sigma (x-y)[\xi
(x)X(x,y)\xi (y)]\\&&X(x,y)=P{\rm
exp}\left[{-i\int_{x}^{y}{\Gamma }_{\mu }(z)d{z}^{\mu
}}\right]\\&&{\Gamma }_{\mu
}={\frac{i}{2}}\left[{\xi ({\partial }_{\mu }-i{R}_{\mu }){\xi
}^{\dagger }+{\xi }^{\dagger }({\partial }_{\mu }-i{L}_{\mu })\xi
}\right]\,\,,\,\,\xi (x)^2\equiv U(x)\\&&{R}_{\mu }={v}_{\mu
}+{a}_{\mu }{\gamma }_{5}+{\frac{1}{2}}s{\gamma }_{\mu
}-{\frac{i}{2}}p{\gamma }_{\mu }{\gamma }_{5},\,\,{L}_{\mu
}={v}_{\mu }-{a}_{\mu }{\gamma }_{\mu }+{\frac{1}{2}}s{\gamma }_{\mu
}-{\frac{i}{2}}p{\gamma }_{\mu }{\gamma }_{5}\end{eqnarray} The
origin of the dependence on $s(x)$ and $p(x)$ is explained in
\cite{lew}.

We have arrived at the GNC quark model, which when expanded in powers
of derivatives and quark mass, successfully models the low energy
chiral Lagrangian of QCD. But our derivation raises a further issue.
When we froze all the massive auxiliary field degrees of freedom, we
not only removed the effect of these fields in loops, but we also
removed their tree-level effects. In the context of a loop expansion,
at lowest order we should keep the effects in the low energy theory
of the tree-level exchange of massive fields. In particular the
original GNC model should be supplemented by the effects of scalar
exchange. This would supplement the quark loop effects, which by
our discussion of duality, are related to the effects of vector and
axial-vector resonances.

\section{Effective Action} In this section we shall present a
derivation of the effective action and the loop expansion without
dropping the cubic and quartic gluon couplings. The auxiliary fields
we introduce will again correspond to nonperturbative amplitudes
only. After integrating out the gauge fields, the generating
functional becomes
\begin{eqnarray} 
Z[J]&=&e^{iW[J]}\nonumber\\
&=&\int{\cal
D}\psi{\cal
D}\overline{\psi}\exp
i{\{}{\int}d^{4}x
\overline{\psi}(i\partial\!\!\!/+J)
+\sum^{\infty}_{n=2}{\int}d^{4}x_1\cdots{d^4}x_{n}
\frac{i^{n}}{n!}
K^{\sigma_1\rho_1\cdots\sigma_n\rho_n}_{\alpha_1\beta_1\cdots\alpha_n\beta_n
}
(x_1,\cdots,x_n)\nonumber\\
&&\times[\overline{\psi}^{i_1\sigma_1}_{\alpha_1}(x_1)
{\psi}^{i_1\rho_1}_{\beta_1}(x_1)]
\cdots[\overline{\psi}^{i_n\sigma_n}_{\alpha_n}(x_n)
{\psi}^{i_n\rho_n}_{\beta_n}(x_n)]
\label{faction}
\end{eqnarray}
See Fig. 1.
\begin{figure}
\begin{center}
\includegraphics{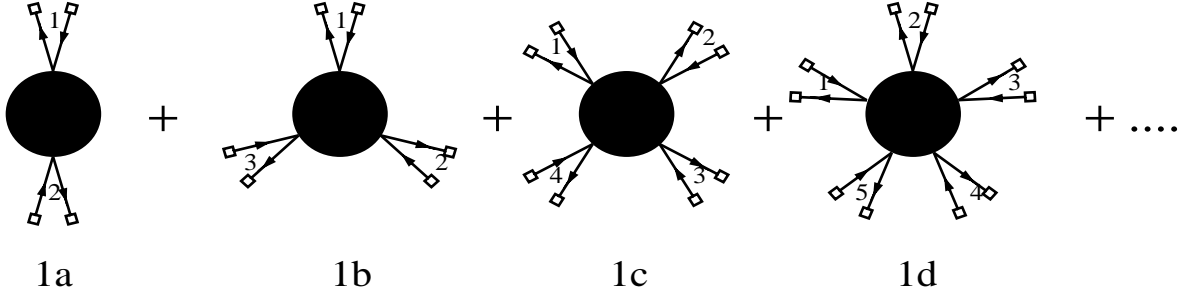}
\end{center}
\caption{Effective interactions caused by integrating out the gauge fields.
The blob with \(2n\) external legs stands for the full gauge boson
\(n\)-point  function without fermion loops. The small empty block stands for
the fermion field.}
\end{figure}
The
$K^{\sigma_1\rho_1\cdots\sigma_n\rho_n}_{\alpha_1\beta_1\cdots\alpha_n\beta_
n}
(x_1,\cdots,x_n)$ are defined
as
follows
\begin{eqnarray}
K^{\sigma_1\rho_1\cdots\sigma_n\rho_n}_{\alpha_1\beta_1\cdots\alpha_n\beta_
n
}
(x_1,\cdots,x_n)
&=&g^{2n}G_{\mu_1\cdots\mu_n}^{a_1\cdots
a_n}(x_1,\cdots,x_n)
(\frac{\lambda_{a_1}}{2})_{\alpha_1\beta_1}(\gamma^{\mu_1})_{\sigma_1\rho_1
}
\cdots
(\frac{\lambda_{a_n}}{n})_{\alpha_n\beta_n}(\gamma^{\mu_n})_{\sigma_n\rho_n
}\nonumber\\
&&+\delta_{n2}g_1^2G_{\mu_1\mu_2}^{(1)}(x_1,x_2)
\delta_{\alpha_1\beta_1}(\gamma^{\mu_1})_{\sigma_1\rho_1}
\delta_{\alpha_2\beta_2}(\gamma^{\mu_2})_{\sigma_2\rho_2}
\end{eqnarray} 
$G^{a_1\cdots a_n}_{\mu_1\cdots\mu_n}(x_1,\cdots,x_n)$ is the full
gluon $n$-point function without fermion loops included and
$G^{(1)}_{\mu\nu}(x,y)$ is the $U(1)$ gauge field propagator without
fermion loops.
 
We may rewrite the sum over $n$ in
(\ref{faction})
as follows.
\begin{eqnarray}
&&\sum^{\infty}_{n=2}{\int}d^{4}x_1\cdots{d^4}x_{n}
\frac{(i)^{n}}{n!}
K^{\sigma_1\rho_1\cdots\sigma_n\rho_n}_{\alpha_1\beta_1\cdots\alpha_n\beta_n
}
(x_1,\cdots,x_n)\nonumber\\
&&\times[\overline{\psi}^{i_1\sigma_1}_{\alpha_1}(x_1)
{\psi}^{i_n\rho_n}_{\beta_n}(x_n)]
[\overline{\psi}^{i_2\sigma_2}_{\alpha_2}(x_2)
{\psi}^{i_1\rho_1}_{\beta_1}(x_1)]
\cdots[\overline{\psi}^{i_n\sigma_n}_{\alpha_n}(x_n)
{\psi}^{i_{n-1}\rho_{n-1}}_{\beta_{n-1}}(x_{n-1})]\nonumber\\
&&=\sum^{\infty}_{n=1}{\int}d^{4}x_1\cdots{d^4}x_{2n}
\frac{(i)^{2n}C_{2n}^n}{(2n)!}
K^{\sigma_1\rho_1\cdots\sigma_{2n}\rho_{2n}}
_{\alpha_1\beta_1\cdots\alpha_{2n}\beta_{2n}}
(x_1,\cdots,x_{2n})\nonumber\\
&&\times[\overline{\psi}^{i_1\sigma_1}_{R,\alpha_1}(x_1)
{\psi}^{i_{2n}\rho_{2n}}_{L,\beta_{2n}}(x_{2n})]
[\overline{\psi}^{i_2\sigma_2}_{L,\alpha_2}(x_2)
{\psi}^{i_1\rho_1}_{R,\beta_1}(x_1)]
\cdots\nonumber\\
&&\cdots[\overline{\psi}^{i_{2n-1}\sigma_{2n-1}}_{R,\alpha_{2n-1}}(x_{2n-1})
{\psi}^{i_{2n-2}\rho_{2n-2}}_{L,\beta_{2n-2}}(x_{2n-2})]
[\overline{\psi}^{i_{2n}\sigma_{2n}}_{L,\alpha_{2n}}(x_{2n})
{\psi}^{i_{2n-1}\rho_{2n-1}}_{R,\beta_{2n-1}}(x_{2n-1})]\nonumber\\
&&+\sum^{\infty}_{n=2}{\int}d^{4}x_1\cdots{d^4}x_{n}
\frac{(i)^{n}}{n!}
\bar{K}^{\sigma_1\rho_1\cdots\sigma_n\rho_n}
_{\alpha_1\beta_1\cdots\alpha_n\beta_n}(x_1,\cdots,x_n)\nonumber\\
&&\times[\overline{\psi}^{i_1\sigma_1}_{\alpha_1}(x_1)
{\psi}^{i_1\rho_1}_{\beta_1}(x_1)]
[\overline{\psi}^{i_2\sigma_2}_{\alpha_2}(x_2)
{\psi}^{i_2\rho_2}_{\beta_2}(x_2)]
\cdots[\overline{\psi}^{i_n\sigma_n}_{\alpha_n}(x_n)
{\psi}^{i_n\rho_n}_{\beta_n}(x_n)]
\end{eqnarray}  
where $\bar{K}$ is defined by this expression and
$C^m_n=\frac{n!}{m!(n-m)!}$. See Fig. 2. 
\begin{figure}
\begin{center}
\includegraphics{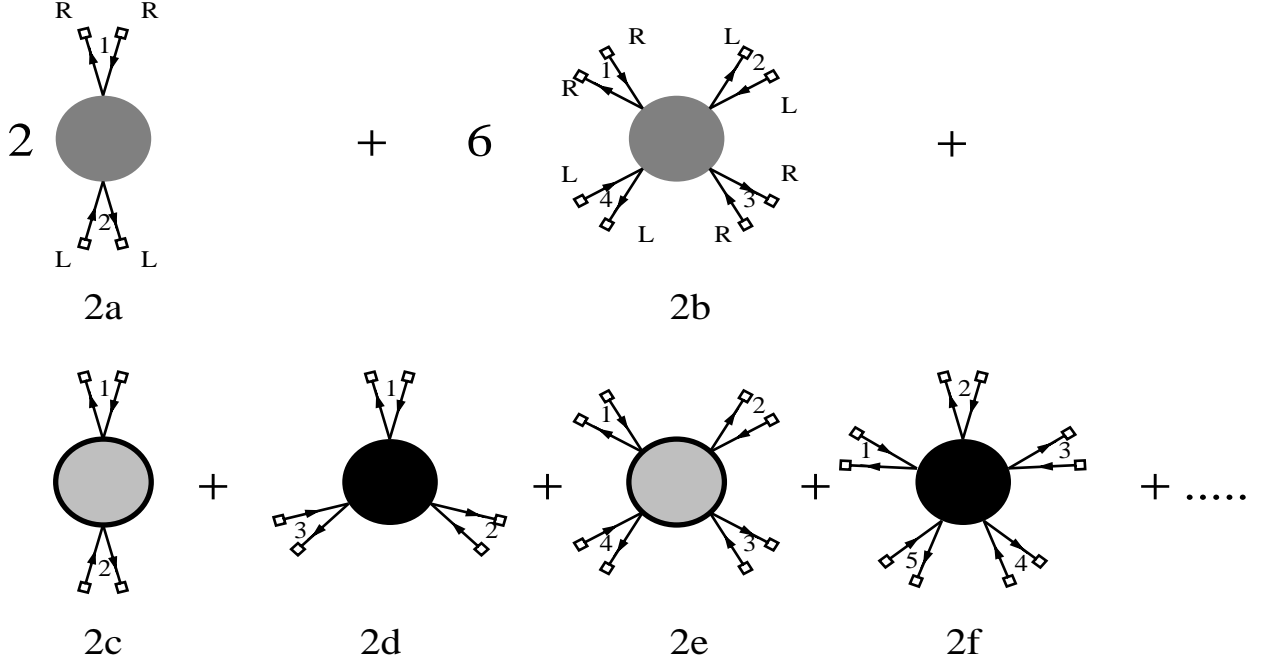}
\end{center}
\caption{The chirality (denoted by L and R) changing part is explicitly
separated out, such that diagram 1a (from Fig. 1)  is separated
into diagram 2a (the chirality changing part) and diagram 2c, and diagram
1c is similarly separated into diagrams 2b and 2e.}
\end{figure}
Inserting this back into the
generating functional and introducing auxiliary fields gives the following.
\begin{eqnarray} 
Z[J]&=&e^{iW[J]}\nonumber\\ &=&\int{\cal D}S_R{\cal
D}S_L{\cal D}\overline{F}{\cal D}F{\cal D}
\overline{B}{\cal D}B{\cal D}\psi{\cal D}\overline{\psi}
\delta\bigg(S^{i\sigma j\rho}_{R,\alpha\beta}(x,y) -
\overline{\psi}^{i\sigma}_{L,\alpha}(x){\psi}^{j\rho}_{R,\beta}(y)\bigg)
\nonumber\\ &&\times\delta\bigg(S^{i\sigma
j\rho}_{L,\alpha\beta}(x,y)
-\overline{\psi}^{i\sigma}_{R,\alpha}(x){\psi}^{j\rho}_{L,\beta}(y)\bigg)
\delta\bigg(\overline{F}_{\alpha\beta}^{i\sigma
j\rho}(x,y)
-
\overline{\psi}^{i\sigma}_{\alpha}(x)\overline{\psi}^{j\rho}_{\beta}(y)\bigg
)
\nonumber\\ &&\times\delta\bigg(F_{\alpha\beta}^{i\sigma j\rho}(x,y)
- \psi^{i\sigma}_{\alpha}(x)\psi^{j\rho}_{\beta}(y)\bigg)
\delta\bigg(\overline{B}_{\alpha_1\alpha_2\alpha_3} ^{i_1\sigma_1
i_2\sigma_2 i_3\sigma_3}(x_1,x_2,x_3) -
\overline{\psi}^{i_1\sigma_1}_{\alpha_1}(x_1)
\overline{\psi}^{i_2\sigma_2}_{\alpha_2}(x_2)
\overline{\psi}^{i_3\sigma_3}_{\alpha_3}(x_3)\bigg)
\nonumber\\ &&\times\delta\bigg(B_{\alpha_1\alpha_2\alpha_3}
^{i_1\sigma_1 i_2\sigma_2 i_3\sigma_3}(x_1,x_2,x_3) -
\psi^{i_1\sigma_1}_{\alpha_1}(x_1)\psi^{i_2\sigma_2}_{\alpha_2}(x_2)
\psi^{i_3\sigma_3}_{\alpha_3}(x_3)\bigg) \exp i{\bigg\{}{\int}d^{4}x
\overline{\psi}(i\partial\!\!\!/+J){\psi}\nonumber\\
&&+\sum^{\infty}_{n=1}{\int}d^{4}x_1\cdots{d^4}x_{2n}
\frac{(i)^{2n}C_{2n}^n}{(2n)!}
K^{\sigma_1\rho_1\cdots\sigma_{2n}\rho_{2n}}
_{\alpha_1\beta_1\cdots\alpha_{2n}\beta_{2n}}
(x_1,\cdots,x_{2n})\nonumber\\ &&\times S^{i_1\sigma_1
i_{2n}\rho_{2n}}_{L,\alpha_1\beta_{2n}}(x_1,x_{2n}) S^{i_2\sigma_2
i_1\rho_1}_{R,\alpha_2\beta_1}(x_2,x_1)
\cdots \nonumber\\ &&\cdots
S^{i_{2n-1}\sigma_{2n-1}i_{2n-2}\rho_{2n-2}}_{L,\alpha_{2n-1}\beta_{2n-2}}
(x_{2n-1},x_{2n-2})
S^{i_{2n}\sigma_{2n}i_{2n-1}\rho_{2n-1}}_{R,\alpha_{2n}\beta_{2n-1}}
(x_{2n},x_{2n-1})\nonumber\\
&&-\sum^{\infty}_{n=1}{\int}d^{4}x_1\cdots{d^4}x_{2n}
\frac{(i)^{2n}}{(2n)!}
\bar{K}^{\sigma_1\rho_1\cdots\sigma_{2n}\rho_{2n}}
_{\alpha_1\beta_1\cdots\alpha_{2n}\beta_{2n}}(x_1,\cdots,x_{2n})\nonumber\\
&&\times
\overline{F}^{i_{2n}\sigma_{2n}i_1\sigma_1}_{\alpha_{2n}\alpha_1}(x_{2n},x_1)
\cdots \overline{F}^{i_{2n-2}\sigma_{2n-2}i_{2n-1}\alpha_{2n-1}}
_{\alpha_{2n-2}\alpha_{2n-1}}(x_{2n-2},x_{2n-1})\nonumber\\ &&\times
F^{i_1\rho_1 i_2\rho_2}_{\beta_1\beta_2}(x_1,x_2)
\cdots F^{i_{2n-1}\rho_{2n-1}i_{2n}\rho_{2n}}
_{\beta_{2n-1}\beta_{2n}}(x_{2n-1},x_{2n})\nonumber\\
&&-\sum^{\infty}_{n=1}{\int}d^{4}x_1\cdots{d^4}x_{2n+1}
\frac{(i)^{2n+1}}{(2n+1)!}
\bar{K}^{\sigma_1\rho_1\cdots\sigma_{2n+1}\rho_{2n+1}}
_{\alpha_1\beta_1\cdots\alpha_{2n+1}\beta_{2n+1}}(x_1,\cdots,x_{2n+1})
\nonumber\\ &&\times\overline{B}^{i_1\sigma_1 i_2\sigma_2 i_3\sigma_3}
_{\alpha_1\alpha_2\alpha_3}(x_1,x_2,x_3)
\overline{F}^{i_4\sigma_4 i_5\sigma_5}_{\alpha_4\alpha_5}(x_4,x_5)
\cdots\overline{F}^{i_{2n}\sigma_{2n} i_{2n+1}\sigma_{2n+1}}
_{\alpha_{2n}\alpha_{2n+1}}(x_{2n},x_{2n+1})\nonumber\\ &&\times
B^{i_1\rho_1 i_2\rho_2 i_3\rho_3}_{\alpha_1\alpha_2\alpha_3}
(x_1,x_2,x_3) F^{i_4\rho_4 i_5\rho_5}_{\alpha_4\alpha_5}(x_4,x_5)
\cdots
F^{i_{2n}\rho_{2n}i_{2n+1}\rho_{2n+1}}_{\alpha_{2n}\alpha_{2n+1}}
(x_{2n},x_{2n+1})\bigg\}
\end{eqnarray} 
See Fig. 3. 
\begin{figure}
\begin{center}
\includegraphics{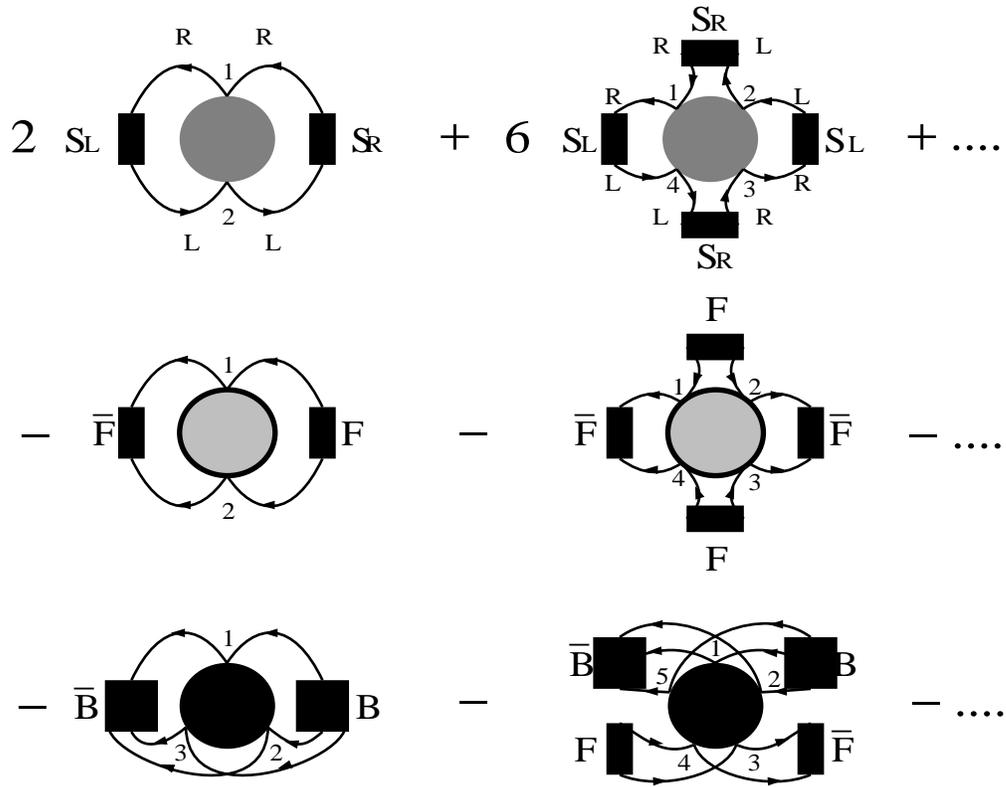}
\end{center}
\caption{The fermion fields in diagrams 2a, 2c, and 2d in Fig. 2 are
combined with the appropriate composite auxiliary fields.}
\end{figure}
We can remove the delta functions by instead including
the following terms in the exponential and integrating over the
additional auxiliary fields, $\chi_L$, $\chi_R$, $U$, $\overline{U}$,
$V$, and $\overline{V}$.
\begin{eqnarray} &&i\int d^4 x d^4 y\bigg[\bigg(S^{i\sigma
j\rho}_{R,\alpha\beta}(x,y) -
\overline{\psi}^{i\sigma}_{L,\alpha}(x){\psi}^{j\rho}_{R,\beta}(y)\bigg)
\chi^{i\sigma j\rho}_{R,\alpha\beta}(x,y)\nonumber\\&& +
\bigg(S^{i\sigma j\rho}_{L,\alpha\beta}(x,y) -
\overline{\psi}^{i\sigma}_{R,\alpha}(x){\psi}^{j\rho}_{L,\alpha}(y)\bigg)
\chi^{i\sigma j\rho}_{L,\alpha\beta}(x,y)
+\bigg(\overline{F}_{\alpha\beta}^{i\sigma
j\rho}(x,y)
-
\overline{\psi}^{i\sigma}_{\alpha}(x)\overline{\psi}^{j\rho}_{\beta}(y)\bigg
)
U_{\alpha\beta}^{i\sigma j\rho}(x,y)\nonumber\\&&
+\overline{U}_{\alpha\beta}^{i\sigma j\rho}(x,y)
\bigg(F_{\alpha\beta}^{i\sigma j\rho}(x,y) -
\psi^{i\sigma}_{\alpha}(x)\psi^{j\rho}_{\beta}(y)\bigg)\bigg]\nonumber\\&&
+i\int d^4 x_1 d^4 x_2 d^4 x_3 \bigg[
\bigg(\overline{B}_{\alpha_1\alpha_2\alpha_3} ^{i_1\sigma_1
i_2\sigma_2 i_3\sigma_3}(x_1,x_2,x_3) -
\overline{\psi}^{i_1\sigma_1}_{\alpha_1}(x_1)
\overline{\psi}^{i_2\sigma_2}_{\alpha_2}(x_2)
\overline{\psi}^{i_3\sigma_3}_{\alpha_3}(x_3)\bigg)
V_{\alpha_1\alpha_2\alpha_3}^{i_1\sigma_1 i_2\sigma_2
i_3\sigma_3}(x_1,x_2,x_3)
\nonumber\\&& +\overline{V}_{\alpha_1\alpha_2\alpha_3} ^{i_1\sigma_1
i_2\sigma_2 i_3\sigma_3}(x_1,x_2,x_3)
\bigg(B_{\alpha_1\alpha_2\alpha_3} ^{i_1\sigma_1 i_2\sigma_2
i_3\sigma_3}(x_1,x_2,x_3) -
\psi^{i_1\sigma_1}_{\alpha_1}(x_1)\psi^{i_2\sigma_2}_{\alpha_2}(x_2)
\psi^{i_3\sigma_3}_{\alpha_3}(x_3)\bigg)\bigg]\nonumber
\end{eqnarray} 
See Fig. 4. 
\begin{figure}
\begin{center}
\includegraphics{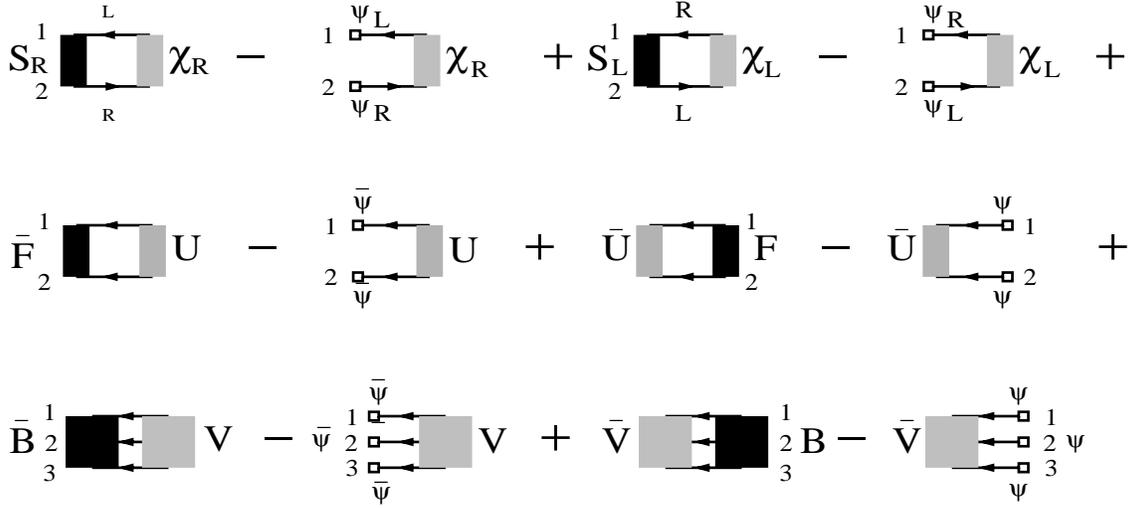}
\end{center}
\caption{These new terms are caused by realizing the constraints
among the fermion fields and the composite auxiliary fields.}
\end{figure}
We may write the result as follows.
\begin{eqnarray} e^{iW[J]}&=&\int{\cal D}S_R{\cal D}S_L{\cal
D}\chi_R{\cal D}\chi_L {\cal D}\overline{F}{\cal D}F{\cal
D}\overline{U}{\cal D}U {\cal D}\overline{B}{\cal D}B{\cal
D}\overline{V}{\cal D}V
 {\rm
exp}i\bigg\{D[\chi_R,\chi_L,\overline{U},U,\overline{V},V,J]\nonumber\\ 
&& +\int d^4 x d^4 y\bigg[ [S^{i\sigma j\rho}_{R,\alpha\beta}(x,y)
\chi^{i\sigma j\rho}_{R,\alpha\beta}(x,y)  + S^{i\sigma
j\rho}_{L,\alpha\beta}(x,y)
\chi^{i\sigma j\rho}_{L,\alpha\beta}(x,y)]
+\overline{F}_{\alpha\beta}^{i\sigma j\rho}(x,y)
U_{\alpha\beta}^{i\sigma j\rho}(x,y)\nonumber\\
&&+\overline{U}_{\alpha\beta}^{i\sigma j\rho}(x,y)
F_{\alpha\beta}^{i\sigma j\rho}(x,y)\bigg]  +\int d^4 x_1 d^4 x_2 d^4
x_3 [
\overline{B}_{\alpha_1\alpha_2\alpha_3} ^{i_1\sigma_1 i_2\sigma_2
i_3\sigma_3}(x_1,x_2,x_3) V_{\alpha_1\alpha_2\alpha_3}^{i_1\sigma_1
i_2\sigma_2 i_3\sigma_3}(x_1,x_2,x_3)
\nonumber\\ &&+\overline{V}_{\alpha_1\alpha_2\alpha_3} ^{i_1\sigma_1
i_2\sigma_2 i_3\sigma_3}(x_1,x_2,x_3) B_{\alpha_1\alpha_2\alpha_3}
^{i_1\sigma_1 i_2\sigma_2 i_3\sigma_3}(x_1,x_2,x_3)]\nonumber\\
&&+\sum^{\infty}_{n=1}{\int}d^{4}x_1\cdots{d^4}x_{2n}
\frac{(i)^{2n}C_{2n}^n}{(2n)!}
K^{\sigma_1\rho_1\cdots\sigma_{2n}\rho_{2n}}
_{\alpha_1\beta_1\cdots\alpha_{2n}\beta_{2n}}
(x_1,\cdots,x_{2n})\nonumber\\ &&\times S^{i_1\sigma_1
i_{2n}\rho_{2n}}_{L,\alpha_1\beta_{2n}}(x_1,x_{2n}) S^{i_2\sigma_2
i_1\rho_1}_{R,\alpha_2\beta_1}(x_2,x_1)
\cdots \nonumber\\ &&\cdots
S^{i_{2n-1}\sigma_{2n-1}i_{2n-2}\rho_{2n-2}}_{L,\alpha_{2n-1}\beta_{2n-2}}
(x_{2n-1},x_{2n-2})
S^{i_{2n}\sigma_{2n}i_{2n-1}\rho_{2n-1}}_{R,\alpha_{2n}\beta_{2n-1}}
(x_{2n},x_{2n-1})\nonumber\\
&&-\sum^{\infty}_{n=1}{\int}d^{4}x_1\cdots{d^4}x_{2n}
\frac{(i)^{2n}}{(2n)!}
\bar{K}^{\sigma_1\rho_1\cdots\sigma_{2n}\rho_{2n}}
_{\alpha_1\beta_1\cdots\alpha_{2n}\beta_{2n}}(x_1,\cdots,x_{2n})\nonumber\\
&&\times
\overline{F}^{i_{2n}\sigma_{2n}i_1\sigma_1}_{\alpha_{2n}\alpha_1}(x_{2n},x_1)
\cdots \overline{F}^{i_{2n-2}\sigma_{2n-2}i_{2n-1}\alpha_{2n-1}}
_{\alpha_{2n-2}\alpha_{2n-1}}(x_{2n-2},x_{2n-1})\nonumber\\ &&\times
F^{i_1\rho_1 i_2\rho_2}_{\beta_1\beta_2}(x_1,x_2)
\cdots F^{i_{2n-1}\rho_{2n-1}i_{2n}\rho_{2n}}
_{\beta_{2n-1}\beta_{2n}}(x_{2n-1},x_{2n})\nonumber\\
&&-\sum^{\infty}_{n=1}{\int}d^{4}x_1\cdots{d^4}x_{2n+1}
\frac{(i)^{2n+1}}{(2n+1)!}
\bar{K}^{\sigma_1\rho_1\cdots\sigma_{2n+1}\rho_{2n+1}}
_{\alpha_1\beta_1\cdots\alpha_{2n+1}\beta_{2n+1}}(x_1,\cdots,x_{2n+1})
\nonumber\\ &&\times\overline{B}^{i_1\sigma_1 i_2\sigma_2 i_3\sigma_3}
_{\alpha_1\alpha_2\alpha_3}(x_1,x_2,x_3)
\overline{F}^{i_4\sigma_4 i_5\sigma_5}_{\alpha_4\alpha_5}(x_4,x_5)
\cdots\overline{F}^{i_{2n}\sigma_{2n} i_{2n+1}\sigma_{2n+1}}
_{\alpha_{2n}\alpha_{2n+1}}(x_{2n},x_{2n+1})\nonumber\\ &&\times
B^{i_1\rho_1 i_2\rho_2 i_3\rho_3}_{\alpha_1\alpha_2\alpha_3}
(x_1,x_2,x_3) F^{i_4\rho_4 i_5\rho_5}_{\alpha_4\alpha_5}(x_4,x_5)
\cdots
F^{i_{2n}\rho_{2n}i_{2n+1}\rho_{2n+1}}_{\alpha_{2n}\alpha_{2n+1}}
(x_{2n},x_{2n+1})\bigg\}\label{main}
\end{eqnarray} 
The path integration over the fermion fields has been isolated.
\begin{eqnarray} && {\rm exp} i
D[\chi_R,\chi_L,\overline{U},U,\overline{V},V,J] =\int{\cal
D}\psi{\cal D}\overline{\psi} {\rm exp}i\bigg\{-\int d^4 x d^4 y\bigg[
\overline{\psi}^{i\sigma}_{L,\alpha}(x){\psi}^{j\rho}_{R,\beta}(y)
\chi^{i\sigma j\rho}_{R,\alpha\beta}(x,y) 
\nonumber\\ && +
\overline{\psi}^{i\sigma}_{R,\alpha}(x){\psi}^{j\rho}_{L,\beta}(y)
\chi^{i\sigma j\rho}_{L,\alpha\beta}(x,y)
+\overline{\psi}^{i\sigma}_{\alpha}(x)\overline{\psi}^{j\rho}_{\beta}(y)
U_{\alpha\beta}^{i\sigma j\rho}(x,y)
+\overline{U}_{\alpha\beta}^{i\sigma j\rho}(x,y)
\psi^{i\sigma}_{\alpha}(x)\psi^{j\rho}_{\beta}(y)\bigg]\nonumber\\
&&-\int d^4 x_1 d^4 x_2 d^4 x_3 \bigg[
\overline{\psi}^{i_1\sigma_1}_{\alpha_1}(x_1)
\overline{\psi}^{i_2\sigma_2}_{\alpha_2}(x_2)
\overline{\psi}^{i_3\sigma_3}_{\alpha_3}(x_3)
V_{\alpha_1\alpha_2\alpha_3}^{i_1\sigma_1 i_2\sigma_2 i_3\sigma_3}
(x_1,x_2,x_3)\nonumber\\ &&+\overline{V}_{\alpha_1\alpha_2\alpha_3}
^{i_1\sigma_1 i_2\sigma_2 i_3\sigma_3}(x_1,x_2,x_3)
\psi^{i_1\sigma_1}_{\alpha_1}(x_1)\psi^{i_2\sigma_2}_{\alpha_2}(x_2)
\psi^{i_3\sigma_3}_{\alpha_3}(x_3)\bigg]
+{\int}d^{4}x\overline{\psi}(i\partial\!\!\!/+J){\psi}\bigg\}
\end{eqnarray}  
                                   
We may use the loop expansion to complete the path integral over the
fields $\chi_R$,
$\chi_L$,$\overline{U}$, $U$, $\overline{V}$ and $V$.
\begin{eqnarray} e^{iW[J]}&=&\int{\cal D}S_R{\cal D}S_L{\cal
D}\overline{F}{\cal D}F {\cal D}\overline{B}{\cal
D}B\exp
i\bigg\{
\Gamma[\chi_{R,c},\chi_{L,c},\overline{U}_c,U_c,\overline{V}_c,V_c,S_R,S_L,
\overline{F},F,\overline{B},B,J]
\nonumber\\
&&+\Gamma_1[\chi_{R,c},\chi_{L,c},\overline{U}_c,U_c,\overline{V}_c,V_c,J]
\bigg\}
\end{eqnarray}
where
$i\Gamma[\chi_R,\chi_L,\overline{U},U,\overline{V},V,S_R,S_L,\overline{F},F,
\overline{B},B,J]$ is the argument of the exponential in (\ref{main})
and
\begin{eqnarray}
&&\Gamma_1[\chi_{R,c},\chi_{L,c},\overline{U}_c,U_c,\overline{V}_c,V_c,J]
\\ &&=\bigg\{\frac{i}{2}{\rm Trln}\bigg[\frac{\partial^2
\Gamma[\phi,\Phi,J]} {\partial
\phi_{\alpha_1\beta_1}^{i_1\sigma_1 j_1\rho_1}(x_1,x'_1)
\partial \phi_{\alpha_2\beta_2}^{i_2\sigma_2
j_2\rho_2}(x_2,x'_2)}\bigg]+
\mbox{all 1PI Feynman vacuum diagrams}\nonumber\\ &&\mbox{ with
inverse of propagator}
\frac{\partial^2 \Gamma[\phi,\Phi,J]} {\partial
\phi_{\alpha_1\beta_1}^{i_1\sigma_1 j_1\rho_1}(x_1,x'_1)
\partial \phi_{\alpha_2,\beta_2}^{i_2\sigma_2 j_2\rho_2}(x_2,x'_2)}
\nonumber\\ &&\mbox{and n-point vertex}
\frac{\partial^n \Gamma[\phi,\Phi,J]} {\partial
\phi_{\alpha_1\beta_1}^{i_1\sigma_1 j_1\rho_1}(x_1,x'_1)\cdots
\partial \phi_{\alpha_n\beta_n}^{i_n\sigma_n j_n\rho_n}(x_n,x'_n)}
\;\;\; n=3,4,....\bigg\}_{\phi=\phi_c}\label{Gamma1}
\end{eqnarray}
$\phi$ is the set of fields $\chi_R$,$\chi_L$,$\overline{U}$,$U$,
$\overline{V}$,$V$, and $\Phi$ is the set of  fields
$S_R$,$S_L$,$\overline{F}$,$F$,$\overline{B}$,$B$.\footnote{
$\overline{V},V,\overline{B},B$ should appear with three indices
instead of two.}
$\Gamma_1[\phi_c,J]$ is independent of $\Phi$ since the
$\Gamma[\phi,\Phi,J]$ in (\ref{Gamma1}) can be replaced by
$D[\phi,J]$ (since $\Gamma[\phi,\Phi,J]-D[\phi,J]$ is linear in
$\phi$).  The classical $\phi_c$ fields satisfy the equations
\begin{equation}
\frac{\partial \{\Gamma[\phi_c,\Phi,J]+ \Gamma_1[\phi_c,J]\}}
{\partial \phi_{c,\alpha\beta}^{i\sigma j\rho}(x,y)}=0\label{phiceq3}
\end{equation} which determines $\phi_c$ as a functional of $\Phi$.
We may again use a loop expansion to integrate out the field $\Phi$.
\begin{equation} W[J]=\Gamma_s[\Phi_c,J]+\Gamma_{s1}[\Phi_c,J]
\end{equation}
\begin{eqnarray}
\Gamma_s[\Phi_c,J]&=&\Gamma[\phi_c,\Phi_c,J]+ \Gamma_1[\phi_c,J]
\label{Gammasdef3}\\
\Gamma_{s1}[\Phi_c,J]&=&\bigg\{
\frac{i}{2}{\rm Trln}\bigg[\frac{\partial^2 \Gamma_s[\Phi,J]}
{\partial \Phi_{\alpha_1\beta_1}^{i_1\sigma_1 j_1\rho_1}(x_1,x'_1)
\partial \Phi_{\alpha_2\beta_2}^{i_2\sigma_2
j_2\rho_2}(x_2,x'_2)}\bigg]+
\mbox{all 1PI Feynman vacuum diagrams}\nonumber\\ &&\mbox{ with
inverse of propagator}
\frac{\partial^2 \Gamma_s[\Phi,J]} {\partial
\Phi_{\alpha_1\beta_1}^{i_1\sigma_1 j_1\rho_1}(x_1,x'_1)
\partial \Phi_{\alpha_2\beta_2}^{i_2\sigma_2
j_2\rho_2}(x_2,x'_2)}\nonumber\\ &&\mbox{and n-point vertex}
\frac{\partial^n \Gamma_s[\Phi,J]} {\partial
\Phi_{\alpha_1\beta_1}^{i_1\sigma_1 j_1\rho_1}(x_1,x'_1)\cdots
\partial \Phi^{i_n\sigma_n j_n\rho_n}_{\alpha_n\beta_n}(x_n,x'_n)}
\;\;\; n=3,4,....\bigg\}_{\Phi=\Phi_c}
\end{eqnarray} The $\Phi_c$ fields satisfy the equations
\begin{equation}
\frac{\partial \{\Gamma_s[\Phi_c,J]+ \Gamma_{s1}[\Phi_c,J]\}}
{\partial \Phi_{c,\alpha\beta}^{i\sigma j\rho}(x,y)}=0
\end{equation} which with the help of (\ref{Gammasdef3}) and
(\ref{phiceq3}) become
\begin{equation}
\frac{\partial \{\Gamma[\phi_c,\Phi_c,J]+ \Gamma_{s1}[\Phi_c,J]\}}
{\partial \Phi_{c,\alpha\beta}^{i\sigma j\rho}(x,y)}=0\label{Phiceq3}
\end{equation} Note how the results in (\ref{phiceq3}) and
(\ref{Phiceq3}) have a similar form.
 
The fields
$\overline{U}$,$U$,$\overline{V}$,$V$,
$\overline{F}$,$F$,$\overline{B}$,$B$ carry nontrivial fermion number
and color,  while fields $\chi_L$,$\chi_R$, $S_L$,$S_R$ and external
sources do not. Because of our dynamical arguments associated with
the most attractive channel we may take the nonvanishing classical
fields, for arbitrary $J$, to be colorless and fermion-number
conserving.
 \begin{eqnarray}
 \Phi_c&=&0\hspace{2cm}\mbox{for } \Phi_c\neq S_L,S_R\\
 \phi_c&=&0\hspace{2cm}\mbox{for } \phi_c\neq \chi_L,\chi_R\\
 S^{i\sigma j\rho}_{^R_L,c,\alpha\beta}(x,y)&=&
 \delta_{\alpha\beta}S^{i\sigma j\rho}_{^R_L}(x,y)\\
 \chi^{i\sigma j\rho}_{^R_L,c,\alpha\beta}(x,y)&=&
 \delta_{\alpha\beta}\chi^{i\sigma j\rho}_{^R_L}(x,y)
 \end{eqnarray} 
The result is that only the first two diagrams in Fig. 3 and and the first four
diagrams in Fig. 4 survive. Then the generating functional is
\begin{eqnarray} 
W[J]&=&-i{\rm Trln}(i\partial\!\!\!/+J-P_L \chi_L
P_L-P_R\chi_R P_R)
\nonumber\\ &&+\int d^4 x d^4 y N_c[S^{i\sigma j\rho}_R(x,y)
\chi^{i\sigma j\rho}_R(x,y) +S^{i\sigma j\rho}_L(x,y)
\chi^{i\sigma j\rho}_L(x,y)]\nonumber\\
&&+W_s[S_R,S_L]+\Gamma_1[\chi_R,\chi_L,J]+\Gamma_{s1}[S_R,S_L,J]\label{W5}
\end{eqnarray} with
\begin{eqnarray} 
W_s[S_R,S_L]&=&
\sum^{\infty}_{n=1}{\int}d^{4}x_1\cdots{d^4}x_{2n}
\frac{(i)^{2n}C_{2n}^n}{(2n)!}
K^{\sigma_1\rho_1\cdots\sigma_{2n}\rho_{2n}}
_{\alpha_1\alpha_2\cdots\alpha_{2n}\alpha_1}
(x_1,\cdots,x_{2n})\nonumber\\ &&\times
S^{i_1\sigma_1 i_{2n}\rho_{2n}}_L(x_1,x_{2n})
S^{i_2\sigma_2 i_1\rho_1}_R(x_2,x_1)
\cdots \nonumber\\ 
&&\cdots S^{i_{2n-1}\sigma_{2n-1}i_{2n-2}\rho_{2n-2}}_L(x_{2n-1},x_{2n-2})
S^{i_{2n}\sigma_{2n}i_{2n-1}\rho_{2n-1}}_R(x_{2n},x_{2n-1})
\label{Wsdef}
\end{eqnarray}
If we apply (\ref{phiceq3}) we find that
\begin{eqnarray}
&&[P_{^R_L}S(y,x)P_{^R_L}]^{j\rho i\sigma}+S^{i\sigma j\rho}_{^R_L}(x,y)
+\frac{1}{N_c}
\frac{\partial \Gamma_1[\chi_R,\chi_L,J]}{\partial
\chi_{^R_L}^{i\sigma j\rho}(x,y)}=0\label{SRLrel}\\
&&-iS^{j\rho i\sigma}(y,x)=
\bigg((i\partial\!\!\!/+J-P_L\chi_L P_L-P_R\chi_R P_R)^{-1}
\bigg)^{j\rho i\sigma}(y,x)\label{Sdef}
\end{eqnarray} 
On the other hand (\ref{Phiceq3}) becomes
\begin{eqnarray}
\chi_{^R_L}^{i\sigma j\rho}(x,y)&=&-\frac{1}{N_c}\bigg[
\frac{\partial W_s[S_R,S_L]}{\partial S_{^R_L}^{i\sigma j\rho}(x,y)}
+\frac{\partial \Gamma_{s1}[S_R,S_L,J]} {\partial
S_{^R_L}^{i\sigma j\rho}(x,y)}\bigg]\label{chiReq}
\end{eqnarray} 
By using the last two equations, 
(\ref{W5}) may be written as
\begin{eqnarray} 
W[J]&=&i{\rm Trln}S+W_s[S_R,S_L]
+\Gamma_1[\chi_R,\chi_L,J]+\Gamma_{s1}[S_R,S_L,J]\nonumber\\
&& -\int d^4 x d^4 y \bigg[S^{i\sigma j\rho}_R(x,y)
\frac{\partial W_s[S_R,S_L]}{\partial S^{i\sigma j\rho}_R(x,y)}
 +S^{i\sigma j\rho}_L(x,y)
\frac{\partial W_s[S_R,S_L]}{\partial
S^{i\sigma j\rho}_L(x,y)}\nonumber\\
&&-S_{R,c}^{i\sigma j\rho}(x,y)\frac{\partial \Gamma_{s1}[S_R,S_L,J]}
{\partial S_{R,c}^{i\sigma j\rho}(x,y)}
-S_{L,c}^{i\sigma j\rho}(x,y)\frac{\partial \Gamma_{s1}[S_R,S_L,J]}
{\partial S_{L,c}^{i\sigma j\rho}(x,y)}\bigg]\label{W7}
\end{eqnarray}  
We may also rewrite (\ref{Sdef}) as
\begin{eqnarray}
\bigg(iS^{-1}-(i\partial\!\!\!/+J)\bigg)^{i\sigma j\rho}(x,y)
&=&\frac{1}{N_c}\bigg[P_R\bigg(
\frac{\partial W_s[S_R,S_L]}{\partial S_R(x,y)}+
\frac{\partial \Gamma_{s1}[S_R,S_L,J]}{\partial S_R(x,y)}\bigg)P_R
\nonumber\\ &&+P_L\bigg(\frac{\partial W_s[S_R,S_L]}{\partial
S_L(x,y)}+
\frac{\partial \Gamma_{s1}[S_R,S_L,J]}{\partial S_L(x,y)}\bigg)P_L
\bigg]^{i\sigma j\rho}\label{SDeq}
\end{eqnarray} 
These are our final full results, with (\ref{SDeq})
playing the role of a generalized SD equation when used along with
(\ref{SRLrel}).
 
If we consider results at lowest order in the loop expansion then 
\begin{eqnarray} 
&&W[J]=i{\rm Trln}S+W_s[S_R,S_L]
\nonumber\\&&\;\;\;\;\;\;\;\;-\int d^4 x d^4 y [S^{i\sigma j\rho}_R(x,y)
\frac{\partial W_s[S_R,S_L]}{\partial S^{i\sigma j\rho}_R(x,y)}
 +S^{i\sigma j\rho}_L(x,y)
\frac{\partial W_s[S_R,S_L]}{\partial S^{i\sigma
j\rho}_L(x,y)}]\label{W8}\\
&&\bigg(iS^{-1}-(i\partial\!\!\!/+J)\bigg)^{i\sigma j\rho}(x,y)
=\frac{1}{N_c}\bigg[P_R
\frac{\partial W_s[S_R,S_L]}{\partial S_R(x,y)}P_R +P_L\frac{\partial
W_s[S_R,S_L]}{\partial S_L(x,y)}P_L
\bigg]^{i\sigma j\rho}\label{SDeq1}\\
&&P_{^R_L}S^{j\rho i\sigma}(y,x)P_{^R_L}
+S^{i\sigma j\rho}_{^R_L}(x,y)=0
\label{SRLrel1}
\end{eqnarray}
We notice that the dependence on $\chi_L$ and $\chi_R$ has been removed.
We could instead express all results in terms of $\chi_L$ and $\chi_R$
by using (\ref{Sdef}).

Alternatively we note the result
(\ref{chiReq})  which now reads
\begin{eqnarray}
\chi_{^R_L}^{i\sigma j\rho}(x,y)&=&-\frac{1}{N_c}
\frac{\partial W_s[S_R,S_L]}{\partial
S_{^R_L}^{i\sigma j\rho}(x,y)}\label{chidef9}
\end{eqnarray} 
This defines the relation between $S$ and $\chi$
fields and suggests a Legendre transform
\begin{equation} 
W_{\chi}[\chi_R,\chi_L]= W_s[S_R,S_L]
+N_c\int
d^4 x d^4 y [S^{i\sigma j\rho}_R(x,y)
\chi^{i\sigma j\rho}_R(x,y) +S^{i\sigma j\rho}_L(x,y)
\chi^{i\sigma j\rho}_L(x,y)]\label{W9}
\end{equation} 
such that $W_{\chi}[\chi_R,\chi_L]$ satisfies
\begin{eqnarray} 
&&\frac{\partial W_{\chi}[\chi_R,\chi_L]}{\partial
\chi_{^R_L}^{i\sigma j\rho}(x,y)} =N_c S_{^R_L}^{i\sigma j\rho}(x,y)
\end{eqnarray} 
We may thus write our generating functional (\ref{W9})
as
\begin{eqnarray} W[J]&=&-i{\rm Trln}(i\partial\!\!\!/+J-P_L \chi_L
P_L-P_R\chi_R P_R) +W_{\chi}[\chi_R,\chi_L]\label{qq}
\end{eqnarray}

To make contact with the last section, if we keep only the lowest
order term in $g$ in $W_s[S_R,S_L]$ then
$W_s[S_R,S_L]$ is bilinear in the fields. As we described in the
introduction, the
neglect of these higher order terms could be justified in the $U(N_c)$, but
not the
$SU(N_c)$ theory. From
(\ref{Wsdef}),
\begin{eqnarray} 
W_s[S_R,S_L]&=&-\frac{N_c}{4}{\int}d^{4}x d^4 y [
\frac{g^2}{2N_c}G_{\mu\nu}^{a
a}(x,y)+g_1^2G_{\mu\nu}^{(1)}(x,y)]
Tr[\gamma^{\mu}S_R^T(x,y)\gamma^{\nu}S_L^T(x,y)]\nonumber\\
&\equiv&-\frac{N_c}{4}{\int}d^{4}x d^4
y{\rm Tr}[S_R^T {\cal K}S_L^T]
\end{eqnarray} 
In this compact notation $\chi_{^R_L}={\cal K}S_{^L_R}^T$, which with
the help of (\ref{Sdef}), (\ref{SRLrel1}), and (\ref{chidef9}) can be
written as
\begin{equation}
\chi_{^R_L}(x,y)
=-i[\frac{g^2}{2N_c}G_{\mu\nu}^{a
a}(x,y)+g_1^2G_{\mu\nu}^{(1)}(x,y)]
\gamma^{\mu}P_{^L_R}(i\partial\!\!\!/+J-P_L\chi_L P_L-P_R\chi_R P_R)^{-1}
P_{^L_R}\gamma^{\mu}
\end{equation} 
This is the SD equation written in terms of the $\chi$ fields. Finally
we have
\begin{eqnarray} 
W_{\chi}[\chi_R,\chi_L]&=&-\frac{N_c}{4}{\int}d^{4}x d^4 y
{\rm Tr}[\chi_L{\cal K}^{-1}\chi_R]
\end{eqnarray} 
With this last result and (\ref{qq}) we essentially
recover the form of the generating functional of the previous
section, with a minor difference being that gluon loop
effects are now included in the gluon propagator.

\section*{Acknowledgement}
 
The work of Q.W. is supported in part by the National Natural Science
Foundation of China and Fundamental Science Research Foundation of
Tsinghua University. The work of B.H. is supported in part by the
Natural Sciences and Engineering Research Council of Canada.


\end{document}